\documentclass[12pt]{article}
\usepackage{amssymb}
\usepackage{amsmath}

\newcommand{\bea}{\begin{eqnarray}}
\newcommand{\eea}{\end{eqnarray}}

\def\fr{\frac}

\def\hw{\hat{w}}

\def\hom{\hat{\omega}}

\def\O{\mathcal{O}}

\def\2{\frac{1}{2}}

\begin{document}
\begin{titlepage}
\begin{flushright}
\begin{tabular}{l}
UTHEP-521\\
hep-th/0604103\\
April 2006
\end{tabular}
\end{flushright}

\vspace{5mm}

\begin{center}
{\Large \bf Bosonic Massless Higher Spin Fields\\
from Matrix Model}
\baselineskip=24pt

\vspace{20mm}
{\large
Takashi Saitou\footnote{e-mail: saitout@het.ph.tsukuba.ac.jp}}\\
{\it Institute of Physics, University of Tsukuba,\\
     Tsukuba, Ibaraki 305-8571, Japan}\\
\vspace{20mm}
\end{center}

\begin{abstract}
We study matrix models as a new approach 
to formulate massless higher spin gauge field theory.
As a first step in this direction, we show that the free 
equation of motion of bosonic massless higher spin gauge fields
can be derived from that of a matrix model. 
\end{abstract}
\end{titlepage}
\newpage


\section{Introduction}
It has been known that there are problems in the construction of 
consistent interactions for 
massless higher spin gauge fields, 
though there are physically acceptable free field Lagrangian 
for them.
Free field Lagrangian for higher spin fields, in which 
these particles are expressed by 
totally symmetric tensor or tensor-spinor 
fields, was originally 
derived by Fronsdal for bosons
\cite{fronsdal} 
and Fang-Fronsdal for 
fermions 
\cite{fangfronsdal}.\footnote{
In four dimensional spacetime, all higher spin fields can be described
either by totally symmetric tensor or totally symmetric tensor-spinor 
fields. This is not the case for a dimension that is larger than four.
In this paper, we will restrict ourselves to the consideration of
symmetric tensor fields.}
There are some approaches for the construction of free field Lagrangian 
\cite{dewitfeerdman}
\cite{vasilievfreehs}.
Interaction problems appear 
when one tries to couple massless higher spin fields to 
an electromagnetic field
\cite{velozwanziger},  
to gravity
\cite{aragonedeser}
\cite{bhwn}
\cite{aragoneroche} 
or to construct self-interactions   
\cite{BB}
\cite{BBV}.
For a review of massless higher spin gauge field theory, see  
\cite{sorokin}. 


At present, there exist various approaches to the theory, 
which  solve the interaction problems in some cases.
For example, an approach, 
called the unfolded formalism, was developed by Vasiliev et
al
\cite{vasilievreview}
\cite{sezginsundell}.
 They succeeded in the construction of interacting higher spin gauge theory 
with a nonzero cosmological constant
\cite{fradkinvasiliev}.
An approach, called BRST approach, was initiated by development of
string field theory on the basis of BRST techniques
\cite{ouvrystern}
\cite{bengtsson}
\cite{spin3selfint}
\cite{buchbinder}
\cite{Sagnotti:2003qa}.

In this paper, we study matrix models as a new approach to formulate 
massless higher spin gauge field theory. 
Recently, it has been shown that the Einstein equation can be obtained
from
the equation of motion of a matrix model by 
introducing a new interpretation of the matrix model, 
in which matrices represent differential operators on a curved spacetime
\cite{matrix}.
Furthermore, it was pointed out that there is a possibility 
that matrix models  
include the degrees of freedom of massless higher spin gauge fields.
An advantage of this formalism is that 
the matrix model possesses a  gauge invariance manifestly, embedded in 
the $U(N)$ symmetry. 
Therefore it is interesting to analyze interacting massless higher spin
gauge field theory using the matrix model.

A first step towards constructing massless higher spin 
gauge field theory is the 
formulation of the free theory. Therefore, 
in this paper we show that the free equation of motion of bosonic 
massless higher spin gauge fields can
      be derived from 
that of the matrix model.

There is another motivation for our study.
Massless higher spin fields 
are expected to appear in the tensionless limit of 
string theory, since mass squared of them are all
proportional to the string tension. Thus in this limit,
one should observe an enhancement of gauge symmetry of string theory 
 by that of
the massless higher spin field theory.
On the other hand, matrix models are expected to be a nonperturbative 
formulation of string theory.
Therefore our study may be useful for better understanding of 
gauge symmetry of string theory and can lead to further understanding 
of nonperturbative aspects of string theory.

\vspace{10pt}

The organization of this paper is as follows.
In section 2, we briefly review the results of  
\cite{matrix}.
In section 3, 
we show that the free equation of motion of 
bosonic massless higher spin gauge fields can be derived
from
that of the matrix model.
Section 4 is devoted to conclusions and future works.


\section{Matrix Model}

In this section, we briefly review the results of
\cite{matrix}.
Introducing a new interpretation of  a matrix model,
 we will see the following three facts.

\begin{itemize}
\item Vacuum Einstein equation can be derived from 
 the equation of motion of the matrix model.

\item There is a possibility that the matrix model can describe 
bosonic massless higher spin fields.

\item Gauge symmetries related to  
higher spin fields are embedded in the $U(N)$ symmetry of the matrix model.

\end{itemize}

\vspace{15pt}

In this paper, 
we consider the large $N$ reduced model of $D$-dimensional pure Yang-Mills
theory
with $U(N)$ gauge symmetry as the matrix model :
\bea
S=-\fr{1}{4g^2}tr \big({[}A_a,A_b{]}{[}A^a,A^b{]}\big), \label{eq:mmaction}
\eea
where $A_a$ are $N \times N$ hermitian bosonic matrices.
Latin indices
denote Euclidean spacetime directions.
This action has the $SO(D)$ Lorentz symmetry 
and $N \times N$ unitary matrix symmetry.
We can also consider supersymmetric version of this model, 
but in this paper we consider only the bosonic action
(\ref{eq:mmaction}).\footnote{
In $D=10$, supersymmetric version of the action (\ref{eq:mmaction}) 
is nothing but the action of IIB matrix model
\cite{ikktmatrix}.}

The basic idea of 
\cite{matrix}
is that matrices represent differential operators on a curved spacetime.
There are several problems with this identification.
For example, matrices act as $Endomorphisms$ on a vector space, which 
means matrices map a vector space to itself. On the contrary, 
covariant derivatives map a tensor field of rank-$n$ to a tensor field 
of rank-$(n+1)$.
Therefore we should prepare a vector space $V$ which contain at least 
tensor fields of any rank and prepare an object $\nabla_{(a)}$ which is 
equivalent to a covariant derivative $\nabla_a$ such that each 
component of $\nabla_{(a)}$ is expressed as an $Endomorphism$ on $V$,
 in order to interpret 
covariant derivatives as matrices. 
In 
\cite{matrix}
the authors showed that $V$ can be given by the space of 
smooth functions on the
principal $Spin(D)$ bundle over a manifold $M$.
A smooth function $f$ on it
is defined as the mapping 
\bea
f : U \times Spin(D) \rightarrow {\mathbb C},
\eea
where $U$ denotes a patch on $M$. Thus, $f$
depends on the local coordinate $(x , g)$, where $x\in U$ and $g\in Spin(D)$, 
of the principal $Spin(D)$ bundle
over $M$.
$\nabla_{(a)}$
can be given by 
\bea
\nabla_{(a)} =R_{(a)}{}^b(g^{-1})\nabla_b ,
\eea
where 
$R_{(a)}{}^b(g)$
 is the vector representation of $Spin(D)$.\footnote{
Though $R_a{}^b(g)$ and $R_{(a)}{}^b(g)$ are the same quantity,
we distinguish them because indices $a$ is transformed by 
the action of $G$, while $(a)$ is not.}
 The covariant derivative  $\nabla_a$ is defined as 
\bea
\nabla_a=e_a{}^\mu(x)(\partial_\mu+\omega_\mu{}^{bc}(x)\O_{bc}),
\eea
where $\O_{bc}(=-\O_{cb})$ is the generator of the 
local Lorentz group $Spin(D)$,
$e_a{}^\mu(x)$
      is the $vierbein$ and $\omega_{\mu,}{}^{ab}(x)$ is the spin-connection.
  Here, Latin indices denote 
the local Lorentz indices. 
Notice that $\nabla_a$ maps a rank-$n$ tensor to a rank-$(n+1)$ tensor and 
$\O_{ab}$ acts on the local Lorentz indices of these tensors. Therefore
we have  
\bea
{[}\O_{ab},\nabla_c{]}=\2(\delta_{ac}\nabla_b-\delta_{bc}\nabla_a),
\eea
in this setting, which will be used later.

Let us see how the Einstein equation can be derived from 
the equation of motion of the matrix model 
by applying  this interpretation to the matrix model.
From the action (\ref{eq:mmaction}), 
we obtain the following equation of motion :
 \bea
{[}A^a,{[}A_a,A_b{]}{]}=0. \label{eq:mmboeq}
\eea
We substitute $A_a=i\nabla_a$ into the equation of motion 
(\ref{eq:mmboeq}), where $i$ is introduced to make $A_a$
hermitian.\footnote{Hermiticity can be 
confirmed by defining the inner product 
$(u,v)=\int ed^dx dg u^*(x,g) v(x,g)$,
where $e=det(e_a{}^\mu)$ , $dg$ is a Haar measure of group $G$,
$u(x,g)$ and $v(x,g)$ are arbitrary functions of $x$ and $g\in G$ } 
The commutator of $A_a$ becomes
\bea
{[}A_a ,A_b {]}
&=&{[}ie_a{}^\mu \nabla_\mu,ie_b{}^\nu \nabla_\nu {]} 
\nonumber \\
&=& -e_a{}^\mu e_b{}^\nu {[}\nabla_\mu,\nabla_\nu {]} 
 -e_{{[}a}{}^\mu (\nabla_\mu e_{b{]}}{}^\nu)\nabla_\nu \nonumber \\
&=& -R_{ab}{}^{cd}\O_{cd} - C^\nu{}_{ab}\nabla_\nu, \label{commutrnabla}
\eea
where $R_{ab,cd}$ is the Riemann tensor 
and $C^\nu{}_{ab}$ is the torsion tensor.
Here, we put the torsionless constraint  $C^\nu{}_{ab}=0$ 
to solve 
the spin-connection in terms of the $vierbein$.
Under the torsionless constraint, the equation of motion becomes
 \bea
{[}A^{a},{[}A_{a},A_{b}{]}{]}
&=&{[}i\nabla^{a},-R_{ab}{}^{cd}{\cal O}_{cd}{]} \cr
&=& -i{[}\nabla^{a},R_{ab}{}^{cd}{]}{\cal O}_{cd}
-iR_{ab}{}^{cd}{[}\nabla^{a},{\cal O}_{cd}{]} \cr
&=&  -i(\nabla^{a}R_{ab}{}^{cd}){\cal O}_{cd}+iR_{b}{}^{c}\nabla_{c}=0,
\label{eq:mmeomnabla}
\eea
and we obtain the following two equations : 
\bea
R_{bc}=0 ,\quad \nabla^{a}R_{ab,cd}=0.  \label{eq:spin2eom}
\eea
The first equation is the vacuum Einstein equation.
The second equation can be derived from the first one and 
the Bianchi identity : 
$\nabla_{a}R^{bc}{}_{de}+\nabla_{b}R^{ca}{}_{de}+\nabla_{c}R^{ab}{}_{de}=0$.
Thus, we have obtained the Einstein equation from the equation of motion of 
the matrix model.

Since each component of matrices $A_a$ 
acts functions on the principal $SO(D)$ bundle over $M$ as an
$Endomorphism$, in general, $A_a$ may be expanded as
\bea
A_a= i\nabla_a +a_a(x)+\fr{i}{2}\{b_a{}^b(x),\nabla_b\}
+\fr{i}{2}\{B_a{}^{bc}(x),\O_{bc}\}
+\fr{i^2}{2}\{e_a{}^{bc}(x),\nabla_b\nabla_c\}+\cdots,
\label{eq:aexpansion}
\eea 
where $i$ and anticommutator $\{\}$ are introduced to make $A_a$
hermitian. The coefficient 
$e_{a,}{}^{bc}(x)$ can be taken to be symmetric under exchange of 
the indices $b\leftrightarrow c$ because antisymmetric part can be absorbed 
in the term that is the first order in $\O_{ab}$. 
Higher order terms expanded in terms of the operators 
$\nabla_a$ and $\O_{ab}$ 
 also can be taken to be symmetric under permutations of the operators. 
We consider the expansion as a sum of homogeneous polynomials 
in $\nabla_a$ and $\O_{ab}$, whose coefficients are identified with 
massless higher spin gauge fields.
Coefficients of the first order homogeneous polynomial express spin-2 
gauge fields, and those of the
 second order one express spin-3 gauge fields and 
so on.
Thus, the number of independent components of higher spin gauge fields 
grows rapidly with degree in $\nabla_a$ and $\O_{ab}$.

Here, we mention how gauge symmetries are embedded in the $U(N)$ symmetry 
of the matrix model. Originally, 
the $U(N)$ symmetry of the matrix model is written as 
\bea
\delta A_a=i {[}\Lambda ,A_a{]}, \label{eq:mmunsym}
\eea
where $\Lambda$ is a $N \times N$ hermitian matrix.
In the new interpretation of the matrix model, 
$\Lambda$ becomes a scalar operator expanded in 
terms of $\nabla_a$ and $\O_{ab}$.

Let us see  how gauge transformations are generated by $\Lambda$ 
in the case of spin-3. In order to deal with this case, we need to 
keep track of only the first and the fifth term in (\ref{eq:aexpansion}),
\bea
A_a = i\nabla_a +(i)^2 \{
e_{a,}{}^{a_1a_2}(x),\nabla_{a_1}\nabla_{a_2}\}+\cdots.
\eea
We take $\Lambda$ as
\bea
\Lambda= (i)^2 \{\lambda^{a_1a_2}(x), \nabla_{a_1}\nabla_{a_2}
\}. 
\eea
Then (\ref{eq:mmunsym}) becomes
\bea
\delta A_a &=& i{[} \Lambda ,A_a {]} \nonumber\\
&=& (\nabla_a\lambda^{a_1a_2}(x))\nabla_{a_1}\nabla_{a_2}+\cdots.
\eea
Thus $e_{a,}{}^{a_1a_2}(x)$ transforms as 
\bea
\delta e_a{}^{a_1a_2}(x)= \nabla_a\lambda^{a_1a_2}(x) +\cdots.
\eea
Therefore, $e_{a,}{}^{a_1a_2}(x)$ transforms as a 
rank-$3$ symmetric tensor field.
Other gauge transformations related to higher spin gauge fields 
are realized in terms of other $\Lambda$.
Thus, 
 this formulation possesses 
the gauge invariance related to higher spin gauge fields 
 manifestly.
This is an advantage of this formulation.


\section{Free Higher Spin Field Equation in Flat Spacetime}

The free 
equation of motion for the rank-$s$ totally symmetric tensor field
$\phi_{\mu_1\cdots\mu_s}$
in $D$-dimensional flat spacetime, 
the so-called ``Fronsdal equation'', is given by 
\bea
W_{\mu_1 \cdots \mu_s}\equiv 
\Box \phi_{\mu_1 \cdots \mu_s}
      -s \partial_{(\mu_1}(\partial \cdot \phi)_{\mu_2 \cdots \mu_s)} 
      +s(s-1)\partial_{(\mu_1}\partial_{\mu_2}\phi'_{\mu_3 \cdots
      \mu_s)}=0  ,
\label{eq:sssymtensoreom}
\eea
where we use notations $(\partial\cdot\phi)_{\mu_1\cdots\mu_{s-1}}
=\partial_\rho\phi^\rho{}_{\mu_1\cdots\mu_{s-1}}$ and 
 $\phi'_{\mu_1\cdots\mu_{s-2}}=\phi^\rho{}_{\rho\mu_1\cdots\mu_{s-2}}$.
Greek indices run from 1 to $D$ and denote flat spacetime directions.
This equation of motion possesses the gauge symmetry, 
 the so-called  Fronsdal symmetry,
\bea
\delta
\phi_{\mu_1\cdots\mu_s}=\partial_{(\mu_1}\lambda_{\mu_2\cdots\mu_s)},
\label{eq:ssgt}
\eea
where the bracket $()$ denotes symmetrization of the flat spacetime 
indices and 
the gauge parameter $\lambda_{\mu_1\cdots\mu_{s-1}}$ is symmetric under 
permutations of the indices.
The conventional formulation for free totally symmetric tensor gauge
fields was originally derived by Fronsdal
\cite{fronsdal}.
The key feature of this formulation is the need for a pair of
constraints, one on the parameter $\lambda_{\mu_1\cdots\mu_{s-1}}$, 
whose trace $\lambda^\rho{}_{\rho\mu_1\cdots\mu_{s-3}}=
\lambda'_{\mu_1\cdots\mu_{s-3}}$ is required to vanish, and one on the
gauge field itself, whose double trace  
$\phi^\rho{}_\rho{}^\sigma{}_{\sigma\mu_1\cdots\mu_{s-5}}=
\phi''_{\mu_1\cdots\mu_{s-5}}$ is also required to vanish. 

In this section, we show that the free equation of motion of 
 higher spin gauge fields (\ref{eq:sssymtensoreom}) 
 can be derived from that of the matrix model (\ref{eq:mmboeq}).
In our formulation, as we will see in the next subsection,
 the constraints on $\phi_{\mu_1\cdots\mu_s}$
 are achieved by putting 
traceless constraints on the fields in $A_a$.
Through the analysis in this section, 
we can understand how  higher spin gauge fields 
are included in the matrix model.

We show the case of spin-3 in subsection 3.1 and  
spin-$s$ in subsection 3.2.


\subsection{Free Spin-3 Field Equation in Flat Spacetime}

Let us first consider the spin-3 case as an example.
The equation of motion of spin-3 field $\phi_{\mu\nu\rho}$ is given by
\bea
W_{\mu\nu\rho}\equiv 
\Box \phi_{\mu\nu\rho}-\partial_\mu (\partial \cdot \phi)_{\nu\rho} 
            -\partial_\nu (\partial \cdot \phi)_{\rho\mu}
            -\partial_\rho (\partial \cdot \phi)_{\mu\nu} \nonumber\\
            +\partial_\mu \partial_\nu \phi'_\rho
            +\partial_\nu \partial_\rho \phi'_\mu
            +\partial_\rho \partial_\mu \phi'_\nu=0. \label{eq:3symtensoreom}
\eea
In this and the next subsection, in order to
derive free field equations we keep only terms linear with 
respect to the component fields and 
we use the ordinary differential operator $\partial_\mu$ instead of 
the covariant derivative $\nabla_a$. 
Notice that the commutation relation of the operators is given by
\bea
{[}\O_{\mu\nu},\partial_\rho{]}
=\2(\delta_{\mu\rho}\partial_\nu-\delta_{\nu\rho}\partial_\mu).
\eea
In order to deal with the spin-3 case, we keep track of the second order 
homogeneous polynomial of the operators $\partial_\mu$ and 
$\O_{\mu\nu}$ in $A_\mu$ : 
\bea
A_\mu=i\partial_\mu +i^2 e_{\mu,}{}^{\nu\rho}(x)\partial_\nu\partial_\rho
    +i^2\2\omega_{\mu,}{}^{\nu,\rho\sigma}(x)
\big( \partial_\nu \O_{\rho\sigma}+\O_{\rho\sigma}\partial_\nu \big)
    \nonumber\\   +i^2\2\Omega_{\mu,}{}^{\rho\sigma,\lambda\kappa}(x)
\big(
\O_{\rho\sigma}\O_{\lambda\kappa}+\O_{\lambda\kappa}\O_{\rho\sigma} 
\big).
\label{eq:spin3exp}
\eea
Here, we do not  write the anticommutator appeared in (\ref{eq:aexpansion}) 
explicitly because terms like $(\partial_\nu
e_\mu{}^{\nu\rho})\partial_\rho$
are  not needed to derive spin-3 field equation.

Based on the analogy of the frame formulation of gravity, 
we can regard that the fields as $e_{\mu,\nu\rho}$ and 
$\omega_{\mu,\nu,\rho\sigma}$ are  
 generalizations of the $vierbein$ and the spin-connection, respectively.
We assume that the fields $e_{\mu,\nu\rho}$
and $\omega_{\mu,\nu,\rho\sigma}$ satisfy the traceless conditions
\bea
e_{\mu,\rho}{}^\rho=0 ,\quad \omega_{\mu,\rho,}{}^\rho{}_\nu=0.
\label{eq:trlscnsteom}
\eea
The totally 
symmetric tensor field $\phi_{\mu\nu\rho}$ is defined in terms of
$e_{\mu,\nu\rho}$ as
\bea
\phi_{\mu\nu\rho}\equiv e_{(\mu,\nu\rho)}=\fr{1}{3}
(e_{\mu,\nu\rho}+e_{\nu,\rho\mu}+e_{\rho,\mu\nu}),
\label{eq:spin3symtensor}
\eea
and $\omega_{\mu,\nu,\rho\sigma}$ and 
      $\Omega_{\mu,\rho\sigma,\lambda\kappa}$ are auxiliary fields.
As we will see shortly, the relation (\ref{eq:spin3symtensor}) 
can be understood from the gauge transformations  
which are embedded in the $U(N)$ symmetry of the matrix model.

Let us summarize the gauge transformations. 
There are three kinds of gauge transformations in the case of spin-3 :
\begin{description}
\item[(i)]  Generalized coordinate transformation, 
generated by 
$\Lambda=\lambda^{\mu\nu}\partial_\mu \partial_\nu$,
\bea
\delta e_{\mu,\nu\rho}=\partial_\mu \lambda_{\nu\rho} ,\quad
   \delta \omega_{\mu,\nu,\rho\sigma}=0 ,\quad
\delta \Omega_{\mu,\rho\sigma,\lambda\kappa}=0, \label{eq:spin3gtd}
\eea
where the  parameter $\lambda_{\mu\nu}$ satisfies 
$\lambda_{\mu\nu}=\lambda_{\nu\mu}$.

\item[(ii)] Generalized local Lorentz transformation, generated by 
$\Lambda=\2 \lambda^{\nu,\rho\sigma}
(\partial_\nu \O_{\rho\sigma}+\O_{\rho\sigma}\partial_\nu)$,
\bea
\delta e_{\mu,\nu\rho}=\lambda_{\nu,\mu\rho}
+\lambda_{\rho,\mu\nu}\equiv \Lambda_{\mu,\nu\rho} ,\quad
   \delta \omega_{\mu,\nu,\rho\sigma}=\partial_\mu \lambda_{\nu,\rho\sigma} 
,\quad \delta \Omega_{\mu,\rho\sigma,\lambda\kappa}=0, \label{eq:spin3gtl}
\eea
where the  parameter $\lambda_{\mu,\nu\rho}$ satisfies 
$\lambda_{\mu,\nu\rho}=-\lambda_{\mu,\rho\nu}$.

\item[(iii)] Auxiliary gauge transformation, generated by 
$\Lambda=\2 \lambda^{\rho\sigma,\lambda\kappa}
(\O_{\rho\sigma}\O_{\lambda\kappa}+\O_{\lambda\kappa}\O_{\rho\sigma})$,
\bea
\delta e_{\mu,\nu\rho}=0,\quad 
   \delta \omega_{\mu,\nu,\rho\sigma}=\lambda_{\mu\nu,\rho\sigma}
+\lambda_{\rho\sigma,\mu\nu},\quad
\delta \Omega_{\mu,\rho\sigma,\lambda\kappa}=\partial_\mu
\lambda_{\rho\sigma,\lambda\kappa}.
\label{eq:spin3gta}
 \eea
where the parameter $\lambda_{\mu\nu,\rho\sigma}$ satisfies 
$\lambda_{\mu\nu,\rho\sigma}=-\lambda_{\nu\mu,\rho\sigma}
  =-\lambda_{\mu\nu,\sigma\rho}$.
\end{description}

Based on the analogy of the frame formulation of gravity, 
the gauge transformation (i), 
which corresponds  to the Fronsdal symmetry, 
is an extension of generalized coordinate
 transformation. 
 The gauge transformation (ii), 
which removes the part that is not totally symmetric in the indices 
 of
 $e_{\mu,\nu\rho}$, is an extension of local Lorentz 
 transformation.
 The gauge symmetry (iii), which does not act
 on dynamical fields, appears only for spins larger than 2.

Under the gauge transformation (i), the rank-3 totally symmetric tensor 
field 
$\phi_{\mu\nu\rho}$ defined in 
(\ref{eq:spin3symtensor}) transforms as follows :
 \bea
\delta \phi_{\mu\nu\rho}
=\partial_\mu\lambda_{\nu\rho}+\partial_\nu\lambda_{\rho\mu}
+\partial_\rho\lambda_{\mu\nu}.
\label{eq:spin3symtensorgtg}
\eea
This is consistent with the identification (\ref{eq:spin3symtensor}).

Since the gauge transformation (i) corresponds to the Fronsdal one, 
in order to derive higher spin field equations we should fix the gauge
symmetries (ii) and (iii). However, gauge symmetries cannot 
remove all extra degrees of freedom.
Therefore, we must impose some constraints that can be used to 
determine the auxiliary fields $\omega_{\mu,\nu,\rho\sigma}$
and $\Omega_{\mu,\rho\sigma,\lambda\kappa}$ in terms of the dynamical field 
$e_{\mu,\nu\rho}$. 
In the rest of this subsection, we will perform these procedures in order.


\vspace{15pt}
\hspace*{-15pt}{\large {\bf  Constraints}}
\vspace{10pt}

Based on the analogy of the frame formulation of gravity, 
we impose  constraints on the field strengths 
in order to solve the auxiliary the fields 
$\omega_{\mu,\nu,\rho\sigma}$ and
$\Omega_{\mu,\rho\sigma,\lambda\kappa}$
 in terms of $e_{\mu,\nu\rho}$.
The field strengths are coefficients of 
the operators in the commutator of $A_\mu$.
The commutator is calculated as follows :
\bea
{[}A_\mu,A_\nu{]}&=& -i\partial_{[\mu}e_{\nu],}{}^{\rho\sigma}
\partial_\rho\partial_\sigma
  -\fr{i}{2}\partial_{[\mu}\omega_{\nu],}{}^{\rho,\sigma\lambda}
\big( \partial_\rho \O_{\sigma\lambda}+\O_{\sigma\lambda}\partial_\rho\big)
\nonumber \\
&& -\fr{i}{2}\Omega_{[\nu,}{}^{\rho\sigma,\lambda\kappa}
{[}\partial_{[\mu},(\O_{\rho\sigma}\O_{\lambda\kappa}+O_{\lambda\kappa}
\O_{\rho\sigma}){]}
\nonumber\\
&& -\fr{i}{2}\partial_{[\mu}\Omega_{\nu],}{}^{\rho\sigma,\lambda\kappa}
(\O_{\rho\sigma}\O_{\lambda\kappa}+O_{\lambda\kappa}\O_{\rho\sigma})
\nonumber\\
&&
-\fr{i}{2}\Omega_{[\nu,}{}^{\rho,\sigma\lambda}
{[}\partial_{\mu]},\partial_\rho\O_{\sigma\lambda}+\O_{\sigma\lambda}
\partial_\rho{]}
\nonumber\\
&=& -i\big( \partial_{[\mu} e_{\nu],}{}^{\rho\sigma}
+\omega_{[\nu,}{}^{\rho,\sigma}{}_{\mu]} 
      \big)\partial_\rho\partial_\sigma \nonumber\\
&&
 -\fr{i}{2}(\partial_{[\mu}\omega_{\nu],}{}^{\rho,\sigma\lambda}
+\Omega_{[\mu,\nu]}{}^{\rho,\sigma\lambda}
+\Omega_{[\mu,}{}^{\sigma\lambda,}{}_{\nu]}{}^\rho)
\big(\partial_\rho\O_{\sigma\lambda}+\O_{\sigma\lambda}\partial_\rho \big)
\nonumber\\
&& -\fr{i}{2}\partial_{[\mu}\Omega_{\nu]}{}^{\rho\sigma,\lambda\kappa}
(\O_{\rho\sigma}\O_{\lambda\kappa}+\O_{\lambda\kappa}\O_{\rho\sigma}),
 \label{eq:commu3}
\eea
where the bracket 
$[ \hspace{2pt} ]$ denotes antisymmetrization of indices.
We impose constraints that the coefficients of $\partial^2$ and $\O^2$ 
are equal to 0 : 
\bea
\partial_{[\mu} e_{\nu],}{}^{\rho\sigma}
+\omega_{[\nu,}{}^{\rho,\sigma}{}_{\mu]} =0 ,
\label{eq:torsionless3}
\\
\partial_{[\mu}\Omega_{\nu],\rho\sigma,\lambda\kappa}=0 .
\label{eq:lomegaconstraint} 
\eea
From (\ref{eq:torsionless3}), 
$\omega_{\mu,\nu,\rho\sigma}$ is solved in terms of 
the first order derivatives in $e_{\mu,\nu\rho}$ :
\bea
\omega_{\mu,\nu,\rho\sigma}&=& \2\partial_\rho(e_{\mu,\nu\sigma}
+e_{\sigma,\nu\mu}) \nonumber\\
&& -\2\partial_\sigma (e_{\rho,\nu\mu}+e_{\mu,\nu\rho}) \nonumber\\
&& +\2\partial_\mu(e_{\sigma,\rho\nu}-e_{\rho,\nu\sigma}). \label{eq:omega}
\eea
This is similar to the  torsionless constraint of gravity. 
The constraint on $\Omega_{\mu,\rho\sigma,\lambda\kappa}$ 
(\ref{eq:lomegaconstraint}) implies that 
$\Omega_{\mu,\rho\sigma,\lambda\kappa}$ can be 
written as a ``pure gauge'' configuration,\footnote{
Precisely speaking, this is not pure gauge because 
$\omega_{\mu,\nu,\rho\sigma}$ is also transformed under (iii). }
\bea
\Omega_{\mu,\rho\sigma,\lambda\kappa}
=\partial_\mu\chi_{\rho\sigma,\lambda\kappa},
\label{eq:puregaugelomega}
\eea
where the parameter $\chi_{\rho\sigma,\lambda\kappa}$ satisfies
$\chi_{\rho\sigma,\lambda\kappa}=-\chi_{\sigma\rho,\lambda\kappa}
= -\chi_{\rho\sigma,\kappa\lambda}$.

Imposing these constraints, we obtain
\bea
{[}A^\mu,{[}A_\mu,A_\nu{]}{]} &=& 
-i\big( \partial_{[\mu} \omega_{\nu],}{}^{\rho,\sigma\mu} 
+\partial_{[\mu}\chi_{\nu]}{}^{\rho,\sigma\mu}
+\partial_{[\mu} \chi^{\mu\sigma,\rho}{}_{\nu]}
\big)\partial_\rho\partial_\sigma \nonumber\\
&&
-\fr{i}{2}\Big{[} \partial^\mu\big( \partial_{[\mu} 
\omega_{\nu],}{}^{\rho,\sigma\lambda}
+\partial_{[\mu}\chi_{\nu]}{}^{\rho,\sigma\lambda}
+\partial_{[\mu}\chi^{\sigma\lambda,}{}_{\nu]}{}^\rho  \big)
\Big{]}\nonumber\\
&& \hspace{80pt} \times
\big( \partial_\rho \O_{\sigma\lambda}+\O_{\sigma\lambda}\partial_\rho\big).
\label{eom3}
\eea
Therefore, we can obtain the following equations of motion :
\bea
\partial_{[\mu} \hom_{\nu],}{}^{\mu,\rho\sigma} 
+\partial_{[\mu}\chi_{\nu]}{}^{(\rho,\sigma)\mu}
+\partial_{[\mu} \chi^{\mu(\sigma,\rho)}{}_{\nu]} =0,
\label{eq:mmceomspin3r}\\
\partial^\mu\big( \partial_{[\mu} 
\omega_{\nu],}{}^{\rho,\sigma\lambda}
+\partial_{[\mu}\chi_{\nu]}{}^{\rho,\sigma\lambda}
+\partial_{[\mu}\chi^{\sigma\lambda,}{}_{\nu]}{}^\rho  \big)=0,
\label{eq:mmceomspin3b}
\eea
where we define $\hom$ as
\bea
{\hom}_{\mu,\nu,\rho\sigma}\equiv \omega_{\mu,(\rho,\sigma)\nu}
      =\2\omega_{\mu,\rho,\sigma\nu}+\2\omega_{\mu,\sigma,\rho\nu} .
\label{eq:omegahat}
\eea
The equation (\ref{eq:mmceomspin3b}) follows from (\ref{eq:mmceomspin3r}). 
Therefore, dynamical field equation of motion is 
(\ref{eq:mmceomspin3r}). 
In order to make the equation (\ref{eq:mmceomspin3r}) to be  
symmetric
under permutations of indices $\nu,\rho,\sigma$ 
we should 
impose the following constraint on $\chi_{\mu\nu,\rho\sigma}$,
\bea
\chi_{\mu\nu,\rho\sigma}=-\fr{1}{3}\omega_{[\mu\nu],\rho\sigma}.
\eea
Imposing this constraint,
we find that the equation (\ref{eq:mmceomspin3r}) is
symmetric under permutations of indices and 
is second order derivatives in $e_{\mu,\nu\rho}$.

\vspace{10pt}
Here, we summarize the constraints we imposed in this subsection 
as follows :
 \begin{itemize}
\item Traceless constraints :
\bea
e_{\mu,\rho}{}^\rho=0 ,\quad \omega_{\mu,\rho,}{}^\rho{}_\nu=0.
\label{eq:trlscnsteom2}
\eea

\item Field strength constraints : 
\bea
\partial_{[\mu} e_{\nu],}{}^{\rho\sigma}
+\omega_{[\nu,}{}^{\rho,\sigma}{}_{\mu]} =0 \label{eq:fscntrntomega},
\\
\partial_{[\mu}\Omega_{\nu],\rho\sigma,\lambda\kappa}=0,
\label{eq:fscntrntoomefa}
\eea
where the constraint (\ref{eq:fscntrntoomefa}) can be solved as
: $\Omega_{\mu,\rho\sigma,\lambda\kappa}
=\partial_\mu\chi_{\rho\sigma,\lambda\kappa}$.
\item  Constraint on $\chi_{\mu\nu,\rho\sigma}$ :
\bea
\chi_{\mu\nu,\rho\sigma}=-\fr{1}{3}\omega_{[\mu\nu],\rho\sigma}.
\label{eq:fscntrntchif}
\eea
\end{itemize}
Imposing these constraints, the auxiliary fields 
$\omega_{\mu,\nu,\rho\sigma}$ and
$\Omega_{\mu,\rho\sigma,\lambda\kappa}$ are expressed in terms of 
the dynamical field $e_{\mu,\nu\rho}$ and only a symmetric part 
remains in the equation
of motion (\ref{eq:mmceomspin3r}).

The constraints imposed on the spin-$3$ fields are 
the traceless constraints (\ref{eq:trlscnsteom2}) and 
the field strength constraints (\ref{eq:fscntrntomega}), 
(\ref{eq:fscntrntoomefa}) and (\ref{eq:fscntrntchif}).
These constraints have been imposed in order to express the equation 
(\ref{eq:mmceomspin3r}) 
in terms of the dynamical field $e_{\mu_1,\mu_2\cdots\mu_s}$.
However, viewed from the matrix model, field strengths should be 
introduced as independent degrees of freedom.
There is a possibility that ``the higher spin field strengths`` 
propagate as asymmetric tensor fields.


\vspace{15pt}

\hspace*{-20pt}{\large {\bf Gauge fixing}}

\vspace{10pt}

So far, we analyzed the elimination of the extra degrees of freedom
by imposing the constraints 
(\ref{eq:trlscnsteom2}), 
(\ref{eq:fscntrntomega}), (\ref{eq:fscntrntoomefa}) and
(\ref{eq:fscntrntchif}).
Combining these constraints
and the equation of motion
(\ref{eq:mmceomspin3r})
 we find that the equation (\ref{eq:mmceomspin3r}) 
is expressed in terms of 
second order derivatives in $e_{\mu,\nu\rho}$ and 
is symmetric under permutations of indices.
However, these constraints cannot eliminate all extra degrees of
freedom.
The Fronsdal equation (\ref{eq:3symtensoreom}) is 
expressed in terms of the rank-$3$ totally symmetric tensor field 
$\phi_{\mu\nu\rho}$, but the equation (\ref{eq:mmceomspin3r})
is expressed in terms of ``the spin-3 $vierbein$'' $e_{\mu,\nu\rho}$,
which have the part that is not totally symmetric.
Last remaining extra degrees of freedom is the part that 
is not totally symmetric in the indices of
$e_{\mu,\nu\rho}$.
Thus, we should eliminate this extra degrees of freedom 
and express dynamical variable in terms of $\phi_{\mu\nu\rho}$ 
, in order to show that the equation
(\ref{eq:mmceomspin3r}) 
coincides with the Fronsdal equation (\ref{eq:3symtensoreom}).
This can be done by fixing the gauge symmetries.
Recall that there are three kinds of the gauge transformations 
(i), (ii) and (iii)
and the gauge transformation (i) corresponds to the Fronsdal gauge 
transformation. 
Therefore it seems that 
we should fix the gauge symmetries (ii) and (iii).
As we see, the gauge symmetries (ii) and (iii) can eliminate
the part that is not totally symmetric in the indices  
of $e_{\mu,\nu\rho}$ and we can express dynamical variable 
in terms of the 
rank-3 totally symmetric tensor field : $\phi_{\mu\nu\rho}$.

First, we fix the gauge symmetry (iii).
Gauge fixing can be done by transforming 
$\hom_{\mu,\nu,\rho\sigma} \rightarrow \hw_{\mu,\nu,\rho\sigma}
=\hom_{\mu,\nu,\rho\sigma}
+\Lambda_{\mu\nu,\rho\sigma}$, choosing
the parameter $\Lambda_{\mu\nu,\rho\sigma}$ as
\bea
\Lambda_{\mu\nu,\rho\sigma}&\equiv& -{\hom}_{\mu,\nu,\rho\sigma}
+\2B_{\mu\nu,\rho\sigma}-\fr{1}{4}
   (B_{\mu\rho,\sigma\nu}+B_{\mu\sigma,\rho\nu}
-B_{\rho\nu,\mu\sigma}-B_{\sigma\nu,\mu\rho}) \nonumber\\
&& -\beta \eta_{\mu\rho}
(B'_{\sigma,\nu}+B'_{\nu,\sigma})
-\beta \eta_{\mu\sigma}(B'_{\rho,\nu}+B'_{\nu,\rho}) \nonumber\\
&& -\beta\eta_{\rho\nu}(B'_{\mu,\sigma}+B'_{\sigma,\mu}) 
-\beta\eta_{\sigma\nu}(B'_{\mu,\rho}+B'_{\rho,\mu}) \nonumber\\
&& +2\beta \eta_{\mu\nu}(B'_{\rho,\sigma}
+B'_{\sigma,\rho})+2\beta\eta_{\rho\sigma}(B'_{\mu,\nu}+B'_{\mu,\nu}) ,
\label{eq:omegahatgta}
\eea
where we define $B_{\mu\nu,\rho\sigma}$, 
 $B'_{\mu,\nu}$ and 
$\beta$ as follows :
\bea
B_{\mu\nu,\rho\sigma}&\equiv & {\hom}_{\mu,\nu,\rho\sigma}
-{\hom}_{\nu,\mu,\rho\sigma} \nonumber\\
&=& \partial_\mu e_{\nu,\rho\sigma}-\partial_\nu e_{\mu,\rho\sigma},
\label{eq:defb}\\
B'_{\mu,\nu} &\equiv & B_{\mu,\rho,}{}^{\rho}{}_{\nu},\label{eq:bprime}\\
\beta &= & \fr{1}{4(D-2)}.\label{eq:beta}
\eea
$\Lambda_{\mu\nu,\rho\sigma}$ satisfies the following properties : 
\bea
&& \Lambda_{\mu\nu,\rho\sigma}
=\Lambda_{\nu\mu,\rho\sigma}=\Lambda_{\mu\nu,\sigma\rho} ,\quad
\Lambda_{\mu(\nu,\rho\sigma)}=0, \label{eq:olambdasym} \\ 
&& \Lambda_{\mu\rho,}{}^\rho{}_\nu=0,
\quad \Lambda_{\mu\nu,\rho}{}^\rho=0. \label{eq:olambdatl}
\eea
$\beta$ is determined by the traceless condition (\ref{eq:olambdatl}). 

Next, we fix the gauge symmetry (ii) by transforming
$e_{\mu,\nu\rho}\rightarrow 
\varepsilon_{\mu,\nu\rho} =e_{\mu,\nu,\rho}+\Lambda_{\mu,\nu\rho}$ with
\bea
\Lambda_{\mu,\nu\rho}\equiv 
-e_{\mu,\nu\rho}
+\phi_{\mu\nu\rho}+\alpha(\eta_{\mu\nu} \phi'_\rho
+\eta_{\mu\rho}\phi'_\nu-2\eta_{\nu\rho}\phi'_\mu).
\label{eq:egtl}
\eea
$\Lambda_{\mu,\nu\rho}$ satisfies the following properties :
\bea
&& \Lambda_{\mu,\nu\rho}=\Lambda_{\mu,\rho\nu},
\quad \Lambda_{(\mu,\nu\rho)}=0,\label{eq:elambdasym}\\
&& \Lambda_{\mu,\rho}{}^\rho=0, \label{eq:elambdatl}
\eea
where $\phi'_\mu\equiv \phi_{\mu\rho}{}^\rho$, $\alpha=\fr{1}{2(D-1)}$.
$\alpha$ is determined by the traceless condition (\ref{eq:elambdatl}).
Carrying out these transformation,
we can remove the part that is not totally symmetric 
in the indices of $e_{\mu,\nu\rho}$ and 
we have $e_{\mu,\nu\rho} =\phi_{\mu\nu\rho}$.
Substituting $\hw_{\mu,\nu,\rho\sigma}$ and
$\varepsilon_{\mu,\nu\rho}$
into the equation (\ref{eq:mmceomspin3r}), 
we can show that the equation (\ref{eq:mmceomspin3r}) coincides with
the Fronsdal equation (\ref{eq:3symtensoreom}).

It is worth noting the relation between our formulation and 
 the unfolded formalism of higher spin 
gauge field theory due to Vasiliev
\cite{vasilievreview}.
Vasiliev constructed 
free field Lagrangian using similar method to the one 
we have employed in this paper
\cite{vasilievfreehs}.
Only difference between the unfolded formalism and our formulation is
the appearance of 
$\Omega_{\mu,\rho\sigma,\lambda\kappa}$
in free theory. 
In the unfolded formalism, 
$\Omega_{\mu,\rho\sigma,\lambda\kappa}$ appears in interacting theories
and contribute to higher derivative interactions.
It is interesting to investigate the relation
between our formulation and the unfolded formalism by analyzing 
higher spin interactions.


\subsection{Free Spin-$s$ Equation in Flat Spacetime}

In this subsection, using the same method as the one we have employed in 
the previous subsection, 
we derive the free equation of motion of the rank-$s$ totally symmetric
tensor field  in 
$D$-dimensional flat spacetime 
(\ref{eq:sssymtensoreom})
from that of the matrix model.

In order to deal with the spin-$s$ case, we keep track of 
the $(s-1)$-th order homogeneous polynomial of the operators $\partial_\mu$ and
$\O_{\mu\nu}$ :
 \bea
 A_\mu = i\partial_\mu
 &+& (i)^{s-1}e_{\mu,}{}^{\mu_1\cdots \mu_{s-1}}\partial_{\mu_1}\cdots
 \partial_{\mu_{s-1}}
 \nonumber\\
  & +& \fr{(i)^{s-1}}{s-1}\omega_{\mu,}{}^{\mu_1\cdots \mu_{s-2},\rho_1 
\sigma_1}
 \{
\partial_{\mu_1}\cdots\partial_{\mu_{s-2}}\O_{\rho_1\sigma_1}
\}\nonumber\\
 & +& \fr{(i)^{s-1}}{(s-1)(s-2)}
 \Omega_{\mu,}{}^{\mu_1\cdots \mu_{s-3},\rho_1\sigma_1,\rho_2\sigma_2}
\{\partial_{\mu_1}\cdots\partial_{\mu_{s-3}}
\O_{\rho_1\sigma_1}\O_{\rho_2\sigma_2}\}
 \nonumber\\ 
 &+&\fr{(i)^{s-1}}{(s-1)(s-2)(s-3)}
\tilde{\Omega}_{(1),\mu,}{}^{\mu_1\cdots\mu_{s-4},
\rho_1\sigma_1,\rho_2\sigma_2,
 \rho_3\sigma_3}\{\partial_{\mu_1}\cdots\partial_{\mu_{s-4}}
 \O_{\rho_1\sigma_1}\O_{\rho_2\sigma_2}\O_{\rho_3\sigma_3}
 \} \nonumber\\
 && \quad\quad \vdots
 \nonumber\\  
 &+& \fr{(i)^{s-1}}{(s-1)!}
 \tilde{\Omega}_{(s-3),  \mu}{}^{\rho_1\sigma_1,\cdots,\rho_{s-1}\sigma_{s-1}}
 \{\O_{\rho_1\sigma_1}\cdots\O_{\rho_{s-1}\sigma_{s-1}}\}.
 \label{eq:spinsaexp}
 \eea 
$e_{\mu,\mu_1\cdots\mu_{s-1}}(x)$ is
the ``spin-s $vierbein$'' and 
 $\omega_{\mu,\mu_1\cdots\mu_{s-2},\rho\sigma}(x)$
is  ``the spin-s connection''.
From the discussion in the previous subsection, 
it seems that $\tilde{\Omega}_{(i)} (i=1,\cdots,s-3)$ 
are not necessary to derive the equation of
motion. 
We set these auxiliary fields to zero :   
$\tilde{\Omega}_1=\cdots=\tilde{\Omega}_{s-3}=0$.
Therefore, $A_\mu$ becomes

\bea
A_\mu = i\partial_\mu
&+& (i)^{s-1}e_{\mu,}{}^{\mu_1\cdots \mu_{s-1}}\partial_{\mu_1}\cdots
\partial_{\mu_{s-1}}
\nonumber\\
& +& \fr{(i)^{s-1}}{s-1}\omega_{\mu,}{}^{\mu_1\cdots \mu_{s-2},\rho_1 \sigma_1}
\{\partial_{\mu_1}\cdots\partial_{\mu_{s-2}}\O_{\rho_1\sigma_1}\}\nonumber\\
& +& \fr{(i)^{s-1}}{(s-1)(s-2)}
\Omega_{ \mu,}{}^{\mu_1\cdots \mu_{s-3},\rho_1\sigma_1,\rho_2\sigma_2}
\{\partial_{\mu_1}\cdots\partial_{\mu_{s-3}}\O_{\rho_1\sigma_1}
\O_{\rho_2\sigma_2}\}.
\eea
Here we assume that the spin-s fields satisfy the 
traceless condition, 
\bea
e_{\mu,\mu_1\cdots\mu_{s-2}\rho}{}^\rho=0,\quad 
\omega_{\mu,\mu_1\cdots \mu_{s-2}\rho,}{}^\rho{}_\sigma=0.
\label{eq:spinstraceless}
\eea
The rank-s symmetric
tensor field $\phi_{\mu_1\cdots\mu_s}$ is  defined as  
$\phi_{\mu_1\cdots\mu_s}=e_{(\mu_1,\mu_2\cdots\mu_s)}$.
$\phi_{\mu_1\cdots\mu_s}$ satisfies the double traceless condition 
$\phi^\rho{}_\rho{}^\sigma{}_{\sigma\mu_5\cdots\mu_s}=0$ as a
consequence of the traceless condition (\ref{eq:spinstraceless}).

Let us summarize the gauge transformations.
In the case of spin-s, there are $s$ kinds of gauge transformation. 
Since auxiliary fields $\tilde{\Omega}_{(i)} (i=1,\cdots,s-3)$
are set to be zero, the following three 
kinds of gauge transformations remain :

\begin{description}

\item[(i)] Generalized coordinate transformation, generated by 
$\Lambda= 
\lambda^{\mu_1\cdots \mu_{s-1}}\partial_{\mu_1}\cdots \partial_{\mu_{s-1}}$,
\bea
\delta 
e_{\mu,\mu_1\cdots \mu_{s-1}}=\partial_\mu\lambda_{\mu_1\cdots \mu_{s-1}} 
,\quad \delta \omega_{\mu,\mu_1 \cdots \mu_{s-2},\mu_{s-1}\nu}=0, 
\quad \delta \Omega_{\mu,\mu_1\cdots
	   \mu_{s-3},\rho\sigma,\lambda\kappa}=0 ,
\label{eq:sgtd}
\eea
where $\lambda_{\mu_1\cdots \mu_{s-1}}$ is symmetric under permutations
	   of the indices $\mu_1,\cdots,\mu_{s-1}$.

\item[(ii)] Generalized local Lorentz transformation, generated by  
$\Lambda=\lambda^{\mu_1\cdots \mu_{s-2},\rho\sigma}
\{\partial_{\mu_1}\cdots \partial_{\mu_{s-2}}\O_{\rho\sigma}
\}$,
\bea
&& \delta e_{\mu,\mu_1\cdots \mu_{s-1}}
=\lambda_{(\mu_1\cdots \mu_{s-2},\mu_{s-1})\mu}
\equiv \Lambda_{\mu,\mu_1\cdots \mu_{s-1}}, \nonumber\\
&& \delta \omega_{\mu,\mu_1\cdots \mu_{s-2},\rho\sigma}=
\partial_\mu \lambda_{\mu_1\cdots \mu_{s-2},\rho\sigma},
\nonumber\\
&&\delta \Omega_{\mu,\mu_1\cdots \mu_{s-3},\rho\sigma,\lambda\kappa}=0,
\label{eq:sgtl}
\eea
where $\lambda_{\mu_1\cdots \mu_{s-2},\rho\sigma}$ is symmetric under
	   permutations of the indices $\mu_1,\cdots,\mu_{s-2}$ and
	   satisfies $\lambda_{\mu_1\cdots \mu_{s-2},\rho\sigma}=
-\lambda_{\mu_1\cdots \mu_{s-2},\sigma\rho}$.

\item[(iii)] Auxiliary gauge transformation, generated by 
$\Lambda=\lambda_{\mu_1 \cdots \mu_{s-3},\rho\sigma,\lambda\kappa}
\{ \partial_{\mu_1}\cdots \partial_{\mu_{s-3}}\O_{\rho\sigma}
\O_{\lambda\kappa}\}$,
\bea
&& \delta e_{\mu,\mu_1 \cdots \mu_{s-1}}=0, \nonumber\\
&& \delta \omega_{\mu,\mu_1 \cdots \mu_{s-2},\rho\sigma}=
\lambda_{(\mu_1 \cdots \mu_{s-3},\mu_{s-2})\mu,\rho\sigma}
+\lambda_{(\mu_1\cdots \mu_{s-3},\rho\sigma,\mu_{s-2})\mu},
\nonumber\\
&& \delta\Omega_{\mu,\mu_1\cdots \mu_{s-3},\rho\sigma,\lambda\kappa}
=\partial_\mu\lambda_{\mu_1\cdots \mu_{s-3},\rho\sigma,\lambda\kappa},
\label{eq:sgta}
\eea
where $\lambda_{\mu_1\cdots \mu_{s-3},\rho\sigma,\lambda\kappa}$ is
	   symmetric under permutations of the indices
	   $\mu_1,\cdots,\mu_{s-3}$ and satisfies
$\lambda_{\mu_1\cdots \mu_{s-3},\rho\sigma,\lambda\kappa}
=-\lambda_{\mu_1\cdots \mu_{s-3},\sigma\rho,\lambda\kappa}
=-\lambda_{\mu_1\cdots \mu_{s-3},\rho\sigma,\kappa\lambda}$.
 \end{description}
Therefore, $\phi_{\mu_1\cdots \mu_s}=e_{(\mu_1,\mu_2\cdots \mu_s)}$ 
are transformed under (i) as
\bea
\delta\phi_{\mu_1\cdots \mu_s}=\partial_{(\mu_1}\lambda_{\mu_2\cdots \mu_s)}.
\eea
This is consistent with the identification
$\phi_{\mu_1\cdots \mu_s}=e_{(\mu_1,\mu_2\cdots \mu_s)}$.
In order to derive the equation of motion
(\ref{eq:sssymtensoreom}) from that of the matrix  model, what we must do is  
to impose constraints on the auxiliary fields and fix the gauge symmetries
(ii) and (iii), as in the spin-3 case.

 
\vspace{15pt}

\hspace*{-15pt}{\large {\bf Constraints}}

\vspace{10pt}

Constraints are imposed on the coefficients of $\partial^{s-1}$ and
$\partial^{s-3}\O^2$ in the commutator of $A_\mu$.
We obtain
\bea
\partial_{[\mu} e_{\nu],}{}^{\mu_1\cdots \mu_{s-1}}
+\omega_{[\nu,}{}^{\mu_1\cdots \mu_{s-2},\mu_{s-1}}{}_{\mu]} =0 ,
\label{eq:strtionless} \\
\partial_{[\mu}\Omega_{\nu],\mu_1\cdots \mu_{s-3},
\rho\sigma,\lambda\kappa}=0.\label{eq:strpuregaugeoomega}
\eea
The constraint (\ref{eq:strpuregaugeoomega}) implies that 
$\Omega_{\mu,\mu_1\cdots \mu_{s-3},\rho\sigma,\lambda\kappa}$ 
can be written as a ``pure gauge'' configuration :
\bea
\Omega_{\mu,\mu_1\cdots \mu_{s-3},\rho\sigma,\lambda\kappa}
=\partial_\mu \chi_{\mu_1\cdots \mu_{s-3},\rho\sigma,\lambda\kappa}.
\eea
We impose constraints on 
$\chi_{\mu_1\cdots\mu_{s-3},\rho\sigma,\lambda\kappa}$ as
\bea
 \chi_{\mu_1\cdots\mu_{s-3},\rho\sigma,\lambda\kappa}=-\fr{1}{s}
\omega_{[\rho,\sigma]\mu_1\cdots\mu_{s-3},\lambda\kappa}.
\label{eq:spinschic}
\eea
Imposing these constraints, the commutators of $A_\mu$ become
\bea
{[}A_\mu,A_\nu{]}=\fr{(i)^{s-1}}{(s-1)}\Big{[}
\partial_{[\mu}\omega_{\nu]}{}^{\mu_1\cdots \mu_{s-2},\rho\sigma}
-\Omega_{[\mu}{}^{\mu_1\cdots \mu_{s-3},\mu_{s-2}}{}_{\nu],}{}^{\rho\sigma}
-\Omega_{[\mu}{}^{\mu_1\cdots \mu_{s-3},\rho\sigma,\mu_{s-2}}{}_{\nu],} \Big{]}
\nonumber\\ \times 
\{\partial_{\mu_1}\cdots\partial_{\mu_{s-2}}\O_{\rho\sigma}\},
\label{commuconstraint}
\eea
and
\bea
{[}A^\mu,{[}A_\mu,A_\nu{]}{]}&=&
 -(i)^s\Big{[}  \partial_{[\mu}
\omega_{\nu]}{}^{\mu_1\cdots \mu_{s-2},\mu_{s-1}\mu }
\nonumber\\ 
&& \quad 
-\Omega_{[\mu}{}^{\mu_1\cdots \mu_{s-3},
\mu_{s-2}}{}_{\nu],}{}^{\mu_{s-1}\mu}
-\Omega_{[\mu}{}^{\mu_1\cdots
\mu_{s-3},\mu_{s-1}\mu,\mu_{s-2}}{}_{\nu]} 
\Big{]} 
 \times \{\partial_{\mu_1}\cdots\partial_{\mu_{s-1}}\}
\nonumber\\ 
&& +\fr{(i)^s}{s-1}\partial^\mu \Big( 
\partial_{[\mu}\omega_{\nu]}{}^{\mu_1\cdots \mu_{s-2},\rho\sigma}
\nonumber\\ 
&& \quad 
-\Omega_{[\mu}{}^{\mu_1\cdots \mu_{s-3},
\mu_{s-2}}{}_{\nu],}{}^{\rho\sigma}
-\Omega_{[\mu}{}^{\mu_1\cdots \mu_{s-3},
\rho\sigma,\mu_{s-2}}{}_{\nu]} \Big{)}
\times \{\partial_{\mu_1}
\cdots\partial_{\mu_{s-2}}\O_{\rho\sigma}\}.
\eea
Therefore we obtain the equations of motion 
\bea
\partial_{[\mu}\hom_{\nu]}{}^{\mu,\mu_1\cdots \mu_{s-2}\mu_{s-1} }
-\Omega_{[\mu}{}^{(\mu_1\cdots \mu_{s-3},
\mu_{s-2}}{}_{\nu],}{}^{\mu_{s-1})\mu}
-\Omega_{[\mu}{}^{(\mu_1\cdots \mu_{s-3},
\mu_{s-1}\mu,\mu_{s-2})}{}_{\nu]}=0,
\label{eq:spinseomricci}
\\
\partial^\mu
\Big( \partial_{[\mu}\omega_{\nu]}{}^{\mu_1\cdots \mu_{s-2},\rho\sigma}
-\Omega_{[\mu}{}^{\mu_1\cdots \mu_{s-3},\mu_{s-2}}{}_{\nu],}{}^{\rho\sigma}
-\Omega_{[\mu}{}^{\mu_1\cdots \mu_{s-3},\rho\sigma,
\mu_{s-2}}{}_{\nu]}\Big) =0, \label{eq:spinseombianchi}
\eea
where $\hom_{\mu,\nu,\mu_1\cdots \mu_{s-1}}$ is defined as
\bea
\hom_{\mu,\nu,\mu_1\cdots \mu_{s-1}} \equiv 
\omega_{\mu,(\mu_1\cdots \mu_{s-2},\mu_{s-1})\nu}.
\eea
The second equation (\ref{eq:spinseombianchi}) 
can be  derived by using the first one (\ref{eq:spinseomricci}). 
Therefore, dynamical field equation is the first one. 
Owing to the constraint (\ref{eq:spinschic}),
the equation of motion (\ref{eq:spinseomricci}) are symmetric
under permutations of indices. 

Here, we summarize the constraints we imposed in this subsection 
as follows :
 \begin{itemize}
\item Traceless constraints :
\bea
e_{\mu,\mu_1\cdots\mu_{s-3}\rho}{}^\rho=0,\quad 
\omega_{\mu,\mu_1\cdots \mu_{s-2}\rho,}{}^\rho{}_\sigma=0.
\label{eq:spinstracelesss}
\eea

\item Field strength constraints : 
\bea
\partial_{[\mu} e_{\nu],}{}^{\mu_1\cdots \mu_{s-1}}
+\omega_{[\nu,}{}^{\mu_1\cdots \mu_{s-2},\mu_{s-1}}{}_{\mu]} =0,
\label{eq:strtionlesss} \\
\partial_{[\mu}\Omega_{\nu],\mu_1\cdots \mu_{s-3},
\rho\sigma,\lambda\kappa}=0.\label{eq:strpuregaugeoomegas}
\eea
The constraint (\ref{eq:strpuregaugeoomegas}) 
can be solved as
\bea
\Omega_{\mu,\mu_1\cdots \mu_{s-3},\rho\sigma,\lambda\kappa}
=\partial_\mu \chi_{\mu_1\cdots \mu_{s-3},\rho\sigma,\lambda\kappa}.
\eea
\item  Constraints on 
$\chi_{\mu_1\cdots\mu_{s-3},\rho\sigma,\lambda\kappa}$ :
\bea
 \chi_{\mu_1\cdots\mu_{s-3},\rho\sigma,\lambda\kappa}=-\fr{1}{s}
\omega_{[\rho,\sigma]\mu_1\cdots\mu_{s-3},\lambda\kappa}.
\label{eq:spinschics}
\eea
\end{itemize}
Imposing these constraints, the auxiliary fields 
$\omega_{\mu,\mu_1\cdots \mu_{s-3},\rho\sigma}$ and
$\Omega_{\mu,\mu_1\cdots \mu_{s-3},\rho\sigma,\lambda\kappa}$
are expressed in terms of 
the dynamical field $e_{\mu,\mu_1\cdots\mu_{s-1}}$
 and only a symmetric part 
remains in the equation
of motion (\ref{eq:spinseomricci}).


\vspace{15pt}

\hspace*{-15pt}{\large {\bf Gauge fixing}}

\vspace{10pt}

First, we fix the gauge symmetry (iii) by transforming 
$\hat{\omega}_{\nu,\mu,\mu_1\cdots \mu_{s-1}}
\rightarrow \hw_{\nu,\mu,\mu_1\cdots \mu_{s-1}}=
\hat{\omega}_{\nu,\mu,\mu_1\cdots \mu_{s-1}}
+ \Lambda_{\nu\mu,\mu_1 \cdots \mu_{s-1}} $, choosing 
the parameter $\Lambda_{\nu\mu,\mu_1 \cdots \mu_{s-1}} $
\bea
\Lambda_{\nu\mu,\mu_1 \cdots \mu_{s-1}} 
 &=& -
\omega_{\nu,\mu,\mu_1\cdots \mu_{s-1}}
+\2 B_{\nu\mu,\mu_1\cdots \mu_{s-1}} \nonumber\\
  &&-\2 \{ B_{\nu \mu_1,\mu \mu_2\cdots \mu_{s-1}}
   +B_{\mu \mu_1,\nu \mu_2 \cdots \mu_{s-1}}\}_{\mu_i} \nonumber\\
&& +\beta_1 \eta_{\nu\mu}\{B'_{\mu_1,\mu_2 \cdots \mu_{s-1}}\}_{\mu_i}   
 \nonumber\\
&& +\beta_2 \{ \eta_{\nu \mu_1}B'_{\mu,\mu_2 \cdots \mu_{s-1}}
   +\eta_{\mu \mu_1}B'_{\nu,\mu_2 \cdots \mu_{s-1}}\}_{\mu_i} \nonumber\\
&& -(\beta_1 +\beta_2)\{ \eta_{\nu \mu_1}B'_{\mu_2,\mu \mu_3 \cdots \mu_{s-1}}
   +\eta_{\mu \mu_1}B'_{\mu_2,\nu \mu_3 \cdots \mu_{s-1}} \}_{\mu_i} 
\nonumber\\
&& -\beta_2\{ \eta_{\mu_1 \mu_2} B'_{\nu,\mu \mu_3 \cdots \mu_{s-1}}
   +\eta_{\mu_1\mu_2}B'_{\mu,\nu \mu_3 \cdots \mu_{s-1}}\}_{\mu_i} \nonumber\\
&& +(\beta_1+2\beta_2)\{ \eta_{\mu_1\mu_2}B'_{\mu_3,\nu\mu \mu_4\cdots 
\mu_{s-1}}\}_{\mu_i},
\label{eq:spinsgta}
\eea
where $\{\}_{\mu_i} $ denotes the symmetrization of indices 
$\mu_i (i=1,\cdots,s-1)$. 
We define $B_{\mu\nu,\mu_1\cdots \mu_{s-1}}$, 
$B'_{\mu_1,\mu_2 \cdots \mu_{s-1}}$,
 $\beta_1$ and  $\beta_2$ 
as follows :
\bea
B_{\mu\nu,\mu_1\cdots \mu_{s-1}}&\equiv& 
     {\hom}_{\mu,\nu,\mu_1 
\cdots \mu_{s-1}}-{\hom}_{\nu,\mu,\mu_1 \cdots \mu_{s-1}} \nonumber\\
    &=& \partial_\mu e_{\nu,\mu_1\cdots \mu_{s-1}}-\partial_\nu 
e_{\mu,\mu_1\cdots \mu_{s-1}} ,
\label{eq:bspins}\\
 B'_{\mu_1,\mu_2 \cdots \mu_{s-1}}&\equiv& 
B_{\mu_1 \rho,}{}^\rho{}_{\mu_2 \cdots \mu_{s-1}},\\
 \beta_1 &=& \fr{s-2}{D-2}, 
\\ \beta_2 &= & \fr{D(s-2)}{2(D-2)(3-s-D)}.
\eea
The parameter $\Lambda_{\nu\mu,\mu_1 \cdots \mu_{s-1}}$ is
symmetric under permutations of the indices $\mu_1,\cdots,\mu_{s-1}$
and satisfies
the following properties :
\bea
\Lambda_{\nu\mu,\mu_1 \cdots \mu_{s-1}}&=&
\Lambda_{\mu\nu,\mu_1 \cdots \mu_{s-1}}, \\
\Lambda_{\nu\mu,\mu_1 \cdots \mu_{s-3}\rho}{}^\rho=0,&&
\Lambda_{\nu \rho,}{}^\rho{}_{\mu_1 \cdots \mu_{s-2}}=0.
\label{eq:spinsolambdatl}
\eea
$\beta_1$ and $\beta_2$ are determined by the traceless conditions 
(\ref{eq:spinsolambdatl}).

Next, we fix the gauge symmetry (ii) by transforming 
$e_{\mu_1,\mu_2\cdots\mu_s} \rightarrow 
\varepsilon_{\mu_1,\mu_2\cdots\mu_s}
=e_{\mu_1,\mu_2\cdots\mu_s} + \Lambda_{\mu_1,\mu_2\cdots\mu_s}
$ with
\bea
\Lambda_{\mu_1,\mu_2 \cdots \mu_s}
= -e_{\mu_1,\mu_2 \cdots \mu_s}
+\phi_{\mu_1 \cdots \mu_s}
              +\alpha \eta_{\mu_1(\mu_2}\phi'_{\mu_3 \cdots \mu_s)}
             -\alpha \eta_{(\mu_2\mu_3}\phi'_{\mu_4\cdots \mu_s)\mu_1}, 
\label{eq:spinsgtl}
\eea
where $\alpha=\fr{(s-1)(s-2)}{2(s+D-4)}$.
      $\Lambda_{\mu_1,\mu_2\cdots \mu_s}$ 
is 
symmetric under permutations of indices $\mu_2,\cdots,\mu_s$ 
and satisfies the traceless condition
\bea
\Lambda_{\mu_1,\mu_2 \cdots \mu_{s-2}\rho}{}^\rho=0.
\eea
Carrying out these transformation,
we can remove the part that is not totally symmetric in the indices of 
$e_{\mu_1,\mu_2\cdots\mu_s}$ and 
we have $e_{\mu_1,\mu_2\cdots\mu_s}=\phi_{\mu_1\cdots\mu_s}$.
Substituting $\hw_{\nu,\mu,\mu_1\cdots \mu_{s-1}}$ and 
$\varepsilon_{\mu_1,\mu_2\cdots\mu_s}$ into the equation 
(\ref{eq:spinseomricci}), we can show that 
(\ref{eq:spinseomricci}) coincides with the Fronsdal equation 
(\ref{eq:sssymtensoreom}).


\section{Conclusions and Future Works}

In this paper, we have shown that the
free equation of motion of bosonic massless higher spin gauge fields 
in $D$-dimensional flat spacetime can be derived from  that of 
the matrix model based on 
the new interpretation of the matrix model.
In order to derive 
higher spin field equations,  
we have done the two things : 1) imposing constraints 2) 
performing gauge fixing
procedures. The results of this paper suggest that bosonic massless 
higher spin fields 
can be embedded in the matrix model. 
This is a first step towards 
construction of interacting massless higher spin gauge field theory
by using the matrix model.
Therefore, the matrix model can be used as a new approach to 
formulate massless higher spin gauge field theory.

There are several things which should be studied further.
One is the  derivation of free fermionic massless higher spin 
gauge field 
equations. Recently, it was  shown that supergravity can be embedded in 
 the supermatrix model 
\cite{matrixsugra}.
There is a possibility that fermionic higher spin fields are 
embedded in the supermatrix model.
Another one is 
to construct the interacting massless higher spin gauge
field theory. As mentioned in Introduction, 
it is difficult to construct interacting theory.
Difficulties associated with the requirement of gauge invariance 
can be overcome by using the matrix model because  
it has gauge invariance manifestly.

\vspace{30pt}
\hspace*{-15pt}{\bf Acknowledgments}

I am grateful to N.Ishibashi and K.Murakami for useful
discussions. I also thank K.Araki, Y.Baba, N.Hatano and  S.Katagiri 
for comments.

\end{document}